\documentclass[12pt]{article}
\usepackage{amsfonts,amsthm,latexsym,amssymb,marvosym}
\usepackage{graphicx,lscape,fancyhdr,array}
\usepackage[arrow,matrix,curve]{xy}\SilentMatrices
\def\xyma{\xymatrix@M.7em}
\def\xymas{\xymatrix@M.1em}
\newcommand{\be}{\begin{equation}}
\newcommand{\ee}{\end{equation}}
\newcommand{\ba}{\begin{eqnarray}}
\newcommand{\ea}{\end{eqnarray}}

\def\d{\delta}

\def\h{\eta}

\def\m{\mu}
\def\n{\nu}

\def\IR{\relax{\rm I\kern-.18em R}}
\def\ZZ{\relax{\hbox{\cmss Z\kern-.4em Z}}}

\begin{document}
\begin{flushright}
DFPD/03/TH/35\\
ULB-TH/03-37\\
\end{flushright}

\vspace{.1cm}

\begin{center} {\Large{\bf Mixed symmetry gauge fields in a flat background}}
\end{center}
\vspace{.1cm}
\begin{center} {\large Xavier Bekaert$^{\sharp}$ and Nicolas Boulanger$^{\flat}$,}
\end{center}

\begin{center}{\sl
$\sharp$ Dipartimento di Fisica, Universit\`a degli Studi di Padova\\
Via F. Marzolo 8, 35131 Padova, Italia\\
$\flat$ Physique Th\'eorique et Math\'ematique, Universit\'e Libre
de Bruxelles\\
C.P. 231, B-1050, Bruxelles, Belgium}
\end{center}
\vspace{.1cm}

\begin{abstract}
We present a list of all inequivalent bosonic covariant free
particle wave equations in a flat spacetime of arbitrary
dimension. The wave functions are assumed to have a finite number
of components. We relate these wave equations to equivalent
versions obtained following other approaches.

\vspace{.1cm}

{\it Talk given by X.B. at the International Workshop on
``Supersymmetries and Quantum Symmetries" (SQS 03), Dubna, Russia,
24-29 July 2003.}
\end{abstract}

\vspace{.3cm}

\section{Elementary particles as irreducible representations of the
          Poincar\'e group}\label{Poincare}

Wigner showed that the rules of quantum mechanics, combined with
the principle of special relativity, imply that the classification
of all possible wave equations describing the evolution of the
states of a free relativistic particle moving in the Minkowski
space ${\mathbb M}^4$ is equivalent to the classification of all
unitary irreducible representations (UIRs) of the Poincar\'e
group\footnote{To deal with double-valued representations, i.e.
fermions, we should actually consider the double covering 
${\mathbb R}^4\ltimes SL(2,\mathbb C)$ of
$ISO(3,1)$.} $ISO(3,1)\equiv {\mathbb R}^4\ltimes SO(3,1)$
\cite{Wigner:1939}.

{\it The classification of all linear relativistic wave equations
in Minkowski spacetime} will be referred to as \textsl{Wigner's
programme}. It was completed in 1939 when, using the method of
induced representations, Wigner showed that the UIRs of $ISO(3,1)$
are characterized by two real numbers: the square of the momentum
$p^2$, and the spin $S$ \cite{Wigner:1939}. Physical
considerations further impose\footnote{We use the ``mostly plus''
signature $(-,+,\cdots,+)\,$ for the metric $\h_{\m\n}\,$.}
$p^2=-m^2\leq 0$ (no tachyon) and $2S\in {\mathbb N}\,$ (discrete
spin).

Subsequently, for every UIR of $ISO(3,1)$, a linear partial
differential equation (PDE) is given, the solutions $\phi$ of
which transform according to that representation. The map
$\phi\,:\;{\mathbb R}^4\longrightarrow V$ stands for the particle
wave function, where the vector space $V$ (over $\mathbb C$)
denotes the representation space for the little group $\ell_4$ of
proper Lorentz transformations that preserve the particle's
four-momentum $p_{\m}\,$. Every UIR of the Poincar\'e group is
determined by a UIR of the little group (acting on the spin part
of the wave function) \cite{Wigner:1939}.

In the process of second quantization, the wave function $\phi$ is
interpreted as a classical field which, in turn, is itself
quantized. It is thus of prime importance to derive the
above-mentioned wave equations from a variational principle. In
order to easily control the Poincar\'e invariance when introducing
interaction terms, it is convenient to start with a free covariant
Lagrangian. Thus, \textit{the determination of a corresponding
covariant Lagrangian for each free particle wave equation in
${\mathbb M}^4$} constitutes the second step in the canonical
field quantization scheme. The latter problem will be referred to
as \textsl{Fierz-Pauli's programme}. This programme was initiated
in 1939 \cite{Fierz:1939ix} and was completed in the seventies by
Singh and Hagen for the massive case ($p^2<0$)
\cite{Singh:1974qz+Singh:1974rc} and by Fang and Fronsdal for the
massless case ($p^2=0$) \cite{Fronsdal:1978rb,Fang:1978wz}.

\section{Higher-spin gauge theories}\label{spin}

In all fundamental field theories known to date, Nature seems to
have limited herself to spins $S\leq 2\,$, although in principle
nothing prevents us from theoretically investigating higher-spin
(i.e. $S>2$) elementary fields since, from a group theoretic point
of view, the Lorentz group $SO(3,1)$ allows representations with
any integer (or half-integer) spin. Incidentally,
\textsl{Fronsdal's programme} consists in {\it introducing
consistent interactions among massless higher-spin fields}
\cite{Fronsdal:1978rb}. This problem was stated in 1978 but still
remains an open mathematical question of field theory.
Numerous preliminary results have recently been obtained (see
\cite{Vasiliev:2001ur} and references therein) which reveal
surprising properties of higher-spin gauge fields.

Fierz-Pauli's programme is a very old problem.  Its generalization to 
arbitrary spacetime dimension $D$ constitutes the first step towards the 
introduction of consistent interactions
among {\it arbitrary} higher-spin fields. In the 80's, string field theory brought 
interest in this direction \cite{Siegel:1986zi,Labastida:1987kw}.
Fierz-Pauli's programme generalization is currently under investigation
\cite{Burdik:2001hj,Bekaert:2002dt,deMedeiros:2002,Zinoviev:2002ye}.

The aim of this talk is restricted to the presentation of {\it an
exhaustive list of inequivalent covariant wave equations for free
relativistic particles moving in a flat background ${\mathbb
M}^D\,$}, which will be referred to as \textsl{Bargmann-Wigner's
theorem} (since it was achieved for $D=4$ by those authors
\cite{Bargmann:1948}). This theorem is itself preliminary to
Fierz-Pauli's programme completion\footnote{Actually this programme has been 
completed in the $OSp(1,1\mid 2)$ formalism \cite{Siegel:1986zi}. We are 
grateful to W. Siegel for calling this fact to our attention.}.

To start with, Wigner's programme is easily generalized to the
Poincar\'e group $ISO(D-1,1)\,$ (see, e.g. \cite{Fush}):

\vspace*{.3cm}

{\bf Lemma} (Wigner's programme) \vspace*{3pt}\newline \noindent
{\it Let $\ell_D\subset SO(D-1,1)$ be a little group corresponding
to $p^2\leq 0$ and $p^\mu\neq 0$. Any UIR of $\ell_D$ with
representation space $V$ provides a UIR of the Poincar\'e group
$ISO(D-1,1)$ the representation space of which is the Hilbert
space $\cal H$ (with $L^2$ norm) of positive energy solutions
$\phi:\,{\mathbb R}^D\longrightarrow V$ of the wave
equation\footnote{Boundary conditions and regularity requirements
should be specified when solving PDEs. In the lemma, we assume
that the ``ket" $\phi \in L^2({\mathbb R}^D)\otimes V$. This
choice is convenient because (a) it provides an obvious norm for
$\cal H$, (b) it selects solutions such that
$\mid\phi(x)\mid\stackrel{\mid x\mid
\rightarrow\infty}{\longrightarrow} 0$, and (c) if we consider
$\phi$ as a temperate distribution (since the ``bra" $\phi\in{\cal
S}^\prime({\mathbb R}^D)\otimes V\,$) then we are always allowed
to convert linear PDEs into algebraic equations by going to the
momentum representation.}}
\begin{equation}(\Box-m^2)\phi=0\,.\label{wav}\end{equation}

If $p^\mu\neq 0$, then the little groups $\ell_D$ are isomorphic
to $SO(D-1)$ for $p^2<0\,$, $ISO(D-2)$ for $p^2=0\,$, and
$SO(D-2,1)$ for $p^2>0\,$. A natural requirement is that the field
$\phi$ should possess a \textit{finite number of components}, i.e.
$\mathrm{dim}(V)<\infty\,$. This removes the unphysical tachyonic
representations with $p^2> 0$ because $SO(D-2,1)$ is non-compact.
The UIRs of $ISO(D-2)$ are induced from those of $SO(D-2)$ and,
since we want a finite-dimensional representation, the non-compact
subgroup ${\mathbb R}^{D-2}$ must act trivially on the wave
functions. Moreover, in order to ensure parity invariance, we are
led to consider finite-dimensional irreducible representations
(irreps) of the orthogonal groups $O(D-\ell)$ with $\ell= 1$ when
$p^2<0$, and $\ell= 2$ when $p^2=0\,$. Therefore, the Hilbert
space $\cal H$ for a massive particle in $\mathbb M^D$ is
isomorphic to the one obtained by a dimensional reduction of a
massless particle in $\mathbb M^{D+1}$. By construction, Fierz-Pauli's 
programme for finite-component fields can thus be restricted, without loss of
generality, to the massless case (see \cite{Aragone:yx} for
completely symmetric fields).

Finite-dimensional irreps of $O(n)$ are characterized by Young
diagrams. For the sake of simplicity, the following discussion
will be limited from now on to single-valued (i.e. tensor)
representations of the orthogonal groups. The space of multilinear
applications from ${\mathbb R}^n\otimes\ldots\otimes{\mathbb R}^n$
to ${\mathbb C}$ is denoted by $T({\mathbb R}^n)\,$. We further
denote by $V^{G}_{\mathbf Y}$ the vector space of tensors in
$T({\mathbb R}^n)$ which are irreducible under $G\subset GL(D,\mathbb R)$ 
and whose symmetry properties are associated with the Young diagram
${\mathbf Y}\,$.

We are interested in fields $\phi$ which have representation space
$V=V^{O(D-\ell)}_{\mathbf Y}$. The case $D=4$ is very particular
in the sense that each tensor irrep of $O(2)$ and $O(3)$ is
equivalent to a completely symmetric tensor irrep (pictured by a
one-row Young diagram with $S$ columns for a spin $S$ particle).
This significant simplification enabled the completion of
Fierz-Pauli's programme in ${\mathbb M}^4\,$. When $D>4\,$, more
complicated Young diagrams (corresponding to ``mixed symmetry"
tensor fields) are generated, the analysis of which requires
appropriate mathematical tools.

\section{Bargmann-Wigner's theorem}
\label{covBW}

Unfortunately, by construction the wave equation (\ref{wav}) is
only covariant under the ``little group" $O(D-\ell)\,$, and not
under the Lorentz group $O(D-1,1)\,$. Consequently, more work is
required in order to obtain a version of Bargmann-Wigner's theorem
for ${\mathbb M}^D\,$. The usual technique consists in considering
a new wave equation for a tensor field $\phi:{\mathbb
R}^D\rightarrow V_{\bf Y}^{GL(D,\mathbb R)}$ which is irreducible
under the general linear group in $D$ dimensions, thereby ensuring
Lorentz covariance. The solution space of (\ref{wav}) is denoted by $\Phi_{\bf
Y}$. The dimension of the representation space $V_{\bf
Y}^{GL(D,\mathbb R)}$ is much bigger than the dimension of $V_{\bf
Y}^{O(D-\ell)}$, which generally implies that extra non-physical
degrees of freedom have been added. In other words, the Hilbert
space ${\cal H}_{\bf Y}$ (see the lemma of section \ref{spin}) is
a strict subspace of $\Phi_{\bf Y}$. Furthermore, the scalar
product of $\Phi_{\bf Y}$ is not positive definite.

\subsection{Massive particle}

The massive case is easy to deal with since it is only necessary
to remove the longitudinal components of the corresponding wave
functions to obtain irreps of $O(D-1)\,$ ($\ell=1\,$). To the
mass-shell equation (\ref{wav}), we must add  (i) the
transversality condition \be\partial\cdot\phi=0\label{trans}\ee
and (ii) the tracelessness of the field $\phi\,$. On-shell, the
field $\phi$ is thus irreducible under the group
$O(D-1,1)\,$: it takes values in the representation space $V_{\bf
Y}^{O(D-1,1)}\subset T({\mathbb R}^D)\,$. The covariant equations
(\ref{wav}) and (\ref{trans}) for a traceless field $\phi :
{\mathbb R}^{D} \rightarrow V_{\bf Y}^{O(D-1,1)}$ take us back to
the representation space ${\cal H}_{\bf Y}$.

\subsection{Massless particle}

In the massless case, the situation is a bit more cumbersome. In
order to have irreps of $O(D-2)\,$ ($\ell=2\,$), it is necessary
to remove the components of the corresponding wave functions lying
along the light-cone directions.

A remedy is to introduce redundancies in the solution space
$\Phi_{\bf Y}$ by resorting to gauge symmetries. In mathematical
terms, one quotients $\Phi_{\bf Y}$ by the gauge orbits, which
leads to the original Hilbert space ${\cal H}_{\bf Y}$ of physical
states (one completely fixes the gauge).
This class of relativistic wave equation is essential because it
should enable the realization of Fierz-Pauli's programme for
arbitrary $D\,$. In the ``metric-like"
\cite{Fronsdal:1978rb,Burdik:2001hj,Labastida:1987kw} and
``frame-like" \cite{Vasiliev:1980} approaches, off-shell trace
conditions are further imposed on the gauge field and the gauge
parameters (in order to avoid the use of auxiliary fields). By
relaxing the orthodox requirement of locality, Francia and
Sagnotti were recently able to forego these trace conditions
\cite{Francia:2002aa} (for mixed symmetry fields, see
\cite{Bekaert:2002dt,deMedeiros:2002,Bekaert:2003az}).

Their field equations were elegantly formulated in terms of the
curvature tensor introduced by de Wit and Freedman \cite{deWit:pe}
in 1980. This tensor is invariant under gauge transformations with
unconstrained gauge parameters. It was already used in 1965 by
Weinberg in his analysis of massless higher-spin fields in
$\mathbb M^4$ \cite{Weinberg:1965rz} (also see the inspiring
pedagogical review on higher-spin fields in the book
\cite{Buchbinder:qv}).

To formulate the theorem we make use of a specific choice of
conventions where sets of antisymmetrized indices are priviledged.
More precisely, we define a {\it multiform ``of spin $S$''} as a
field $K:{\mathbb R}^D\rightarrow {\Lambda}_{[S]}({\mathbb R}^D)$
which takes value in the algebra $\Lambda_{[S]}({\mathbb
R}^D)\equiv\otimes^S\Lambda({\mathbb R}^D)$ of polynomials in the
generators $d_i x^\mu$ ($i=1,2,\ldots,S\,;\,
\m=0,1,\ldots,D-1\,$). For fixed $i$, the $d_i x^\mu$'s generate
the exterior algebra $\Lambda({\mathbb R}^D)\,$. The nilpotent
operators $d_i\equiv d_i x^\m\partial_\m$ generalize the usual
exterior differential $d$ of the de Rham complex $\Omega({\mathbb
R}^D)\,$. With the help of the Minkowski metric, we define the
Hodge operators $*_i$ as well as the codifferentials $\d_i\equiv
\,*_i\, d_i\, *_i\,$. A multiform $K$ of spin $S$ is said to be
{\it harmonic} if it is closed ($d_i K=0$) and coclosed ($\d_i
K=0$) for all $i\in\{1,\ldots ,S\}\,$.

\vspace*{.3cm}

{\bf Proposition} (Bargmann-Wigner's theorem)\vspace*{3pt}
\newline \noindent {\it Let
${\mathbf{\overline{Y}}}$ be a Young diagram with at least two
rows of equal length $S$ and let ${\mathbf Y}$ be the Young
diagram obtained by removing the first row of
$\mathbf{\overline{Y}}\,$.

\noindent Any tensor irrep of $O(D-1,1)$ with representation space
$V^{O(D-1,1)}_{\mathbf{\overline{Y}}}$ provides a massless UIR of
the group $IO(D-1,1)$ associated with the Young diagram ${\mathbf
Y}$ the representation space of which is the space of harmonic
irreducible multiforms $K:\,{\mathbb R}^D\longrightarrow
V^{O(D-1,1)}_{\mathbf{\overline{Y}}}$ of spin $S\,$. This latter
space is isomorphic to the Hilbert space ${\cal H}_{\mathbf Y}$ of
physical states $\phi:\,{\mathbb R}^D\longrightarrow
V^{O(D-2)}_{\mathbf{Y}}\,$ that are solutions of (\ref{wav}).}

\vspace*{.5cm}

The proof of the Bargmann-Wigner theorem is straightforward in
both cases ($\ell=1,2$). The corresponding set of differential
equations is equivalent to a set of algebraic conditions on the
components of the Fourier transform of the corresponding tensor
field. An explicit check shows that these algebraic conditions
constrain the tensor components to belong to the appropriate space
$V_{\mathbf Y}^{O(D-\ell)}\,$. Details are given in \cite{Nick}.

The gauge-invariant curvature tensor for completely symmetric
fields \cite{deWit:pe,Weinberg:1965rz} is generalized for a mixed
symmetry gauge field $\phi:\,{\mathbb R}^D\longrightarrow
V^{GL(D,{\mathbb R })}_{\mathbf Y}$ as follows
\cite{Bekaert:2002dt,deMedeiros:2002,Bekaert:2002cz}\be K\equiv
d_1d_2\ldots d_S\phi:\,{\mathbb R}^D\longrightarrow
V^{GL(D,{\mathbb R })}_{\mathbf{\overline{Y}}}\,.\label{curv}\ee
In \cite{Bekaert:2002dt,Bekaert:2002cz}, field equations were
proposed for gauge fields irreducible under $GL(D,{\mathbb R })$.
This was motivated by a systematic generalization of the work
\cite{Hull:2001iu}. On-shell, the curvature tensor $K$ was taken
to be traceless and harmonic, which is a simple generalization of
the Maxwell equations ($S=1$), the linearized Einstein equation
($S=2$) and the Bargmann-Wigner equations\footnote{Indeed,  
Bargmann and Wigner used fields which transform according to the 
$[(S,0)\oplus (0,S)]$ representations of $SL(2,\mathbb C)\,$. In other
words, they considered fieldstrengths \cite{Buchbinder:qv}. 
To compare with the 
works \cite{Lopatin:hz} one must consider their field
equations (i) in the limit where the cosmological constant goes to
zero and (ii) in the gauge where the (frame-like) gauge field is
equal to the (metric-like) completely symmetric field.} of
\cite{Bargmann:1948,Weinberg:1965rz,Lopatin:hz} for completely symmetric gauge
fields of arbitrary spin $S\,$. 
The field equations of \cite{Bekaert:2002dt,Bekaert:2002cz} should be 
equivalent to the ones proposed in \cite{Siegel:1986zi}.
Thanks to our proposition, the
on-shell fieldstrength $K:{\mathbb R}^D\rightarrow
V^{O(D-1,1)}_{\mathbf{\overline{Y}}}$ provides a UIR of the
Poincar\'e group corresponding to the gauge field $\phi:{\mathbb
R}^D\rightarrow V^{O(D-2)}_{\mathbf Y}$ in the light-cone gauge.
Indeed, gauge fields in the light-cone gauge are essentially
fieldstrengths \cite{Siegel:1986zi,Buchbinder:qv}.

The local wave equations that we provide contain many derivatives
of the gauge field because they are built out of the curvature tensor (\ref{curv}),
but nevertheless can simply be brought back to a second order form
upon partial gauge-fixing \cite{Bekaert:2003az}. Our proposition
proves that an apparently ill-behaved higher-derivative field
equation can in fact be the correct one (leaving aside the subtle
issue of a well-behaved realization of Fierz-Pauli's programme).
Following the method sketched in \cite{Bekaert:2003az}, we checked
that the generalized Poincar\'e lemma of \cite{Bekaert:2002dt}
relates the previous higher-derivative field equations to the
local second-order field equations of \cite{Labastida:1987kw} for
any mixed symmetry field. This procedure introduces a
supplementary non-local term which enables to abandon the trace
conditions on the gauge parameters of the local approaches (in
perfect agreement with the results of \cite{Francia:2002aa} for
completely symmetric fields). These non-local second-order field
equations are equivalent to the ones of \cite{deMedeiros:2002}.
This proves the complete generality of the procedure sketched in
\cite{Bekaert:2003az}.

\section*{Acknowledgements}
%
X.B. thanks P. Marchetti for an enlightening discussion on
convenient function spaces to work with, and D. Tonei for her
improvements of the text.

Work supported in part by the ``Actions de Recherche
Concert\'ees", a ``P\^ole d'Attraction Interuniversitaire"
(Belgium), by IISN-Belgium (convention 4.4505.86) and by the
European Commission RTN programme HPRN-CT-2000-00131.
%


\end{document}